\documentclass[aip,reprint,apl]{revtex4-1}
\usepackage{graphicx,graphicx,epsfig}

\usepackage{graphicx}
\usepackage{bm}

\begin{document}
\title{Two-Beam Spin Noise Spectroscopy}

\author{Yuriy V. Pershin}
\email{pershin@physics.sc.edu}
\affiliation{Department of Physics and Astronomy and USC
Nanocenter, University of South Carolina, Columbia, SC 29208, USA}

\author{Valeriy A. Slipko}
\affiliation{Department of Physics and Astronomy and USC
Nanocenter, University of South Carolina, Columbia, SC 29208, USA}
\affiliation{ Department of Physics and Technology, V. N. Karazin
Kharkov National University, Kharkov 61077, Ukraine }

\author{Dibyendu Roy}
\affiliation{Theoretical Division and the Center for Nonlinear Studies, Los Alamos National Laboratory, B213, Los Alamos, NM 87545}

\author{Nikolai A. Sinitsyn}
\affiliation{Theoretical Division, Los Alamos National Laboratory, B213, Los Alamos,
NM 87545}

\begin{abstract}
We propose a method of two-beam spin noise spectroscopy to test the spin transport at equilibrium via analysis of correlations between time-shifted spin fluctuations at different space locations. This method allows one to determine the strength of spin-orbit interaction and spin relaxation time and separate spin noise of conducting electrons from the background noise of localized electrons. We formulate a theory of two-beam spin noise spectroscopy in semiconductor wires with Bychkov-Rashba spin-orbit interaction taking into account several possible spin relaxation channels and finite size of laser beams. Our theory predicts a peak shift with respect to the Larmor frequency to higher or lower frequencies depending on the strength of spin orbit interaction and distance between the beams. The two-beam spin noise spectroscopy could find applications in experimental studies of semiconductors, emergent materials and many other systems.
\end{abstract}


\maketitle

The ability to understand and control the electron spin polarization in semiconductors will have profound technological impacts~\cite{Zutic04a,Bandyopadhyay08a}. For this purpose, it is important to obtain information about interactions that spins of conducting electrons experience in semiconductors as these interactions can play a significant role in the processes of spin relaxation and control. Experimentally, the electron spin polarization and its fluctuations (spin noise) in semiconductors are often detected using the optical Faraday or Kerr rotation technique (see, e.g., Refs. \onlinecite{Young2002a,Muller08a,Muller10a}).
The Faraday rotation technique relies on the fact that a linearly polarized light propagating through a material with a net magnetic moment is subject to a rotation of its plane of polarization. The amount of rotation induced is proportional to the internal local magnetization of material and thus can provide information about electron spin polarization and its fluctuations.

The major advantage of spin noise spectroscopy~\cite{Oestreich2005a,Muller08a,Crooker09a,Muller10a,Glazov2011a} to all
other probes of semiconductor spin dynamics (e.g., optical pump-probe technique~\cite{Furis07a} or methods based on charge transport~\cite{Mishchenko03a,Lamacraft04a,Egues02a,Egues05a}) lies in the fact that in principle no energy has to be dissipated in the sample, i.e., it exclusively yields the
intrinsic, undisturbed spin dynamics \cite{Muller10a}. In addition, the spin noise spectroscopy allows accumulation of a large statistics that smoothes out the statistical noise in the data. At the present time, experimentalists utilize a single beam setup measuring a noise correlation function at the same space location~\cite{Oestreich2005a,Muller08a,Crooker09a,Muller10a}.

\begin{figure}[b]
\centering\includegraphics[width=.85\linewidth]{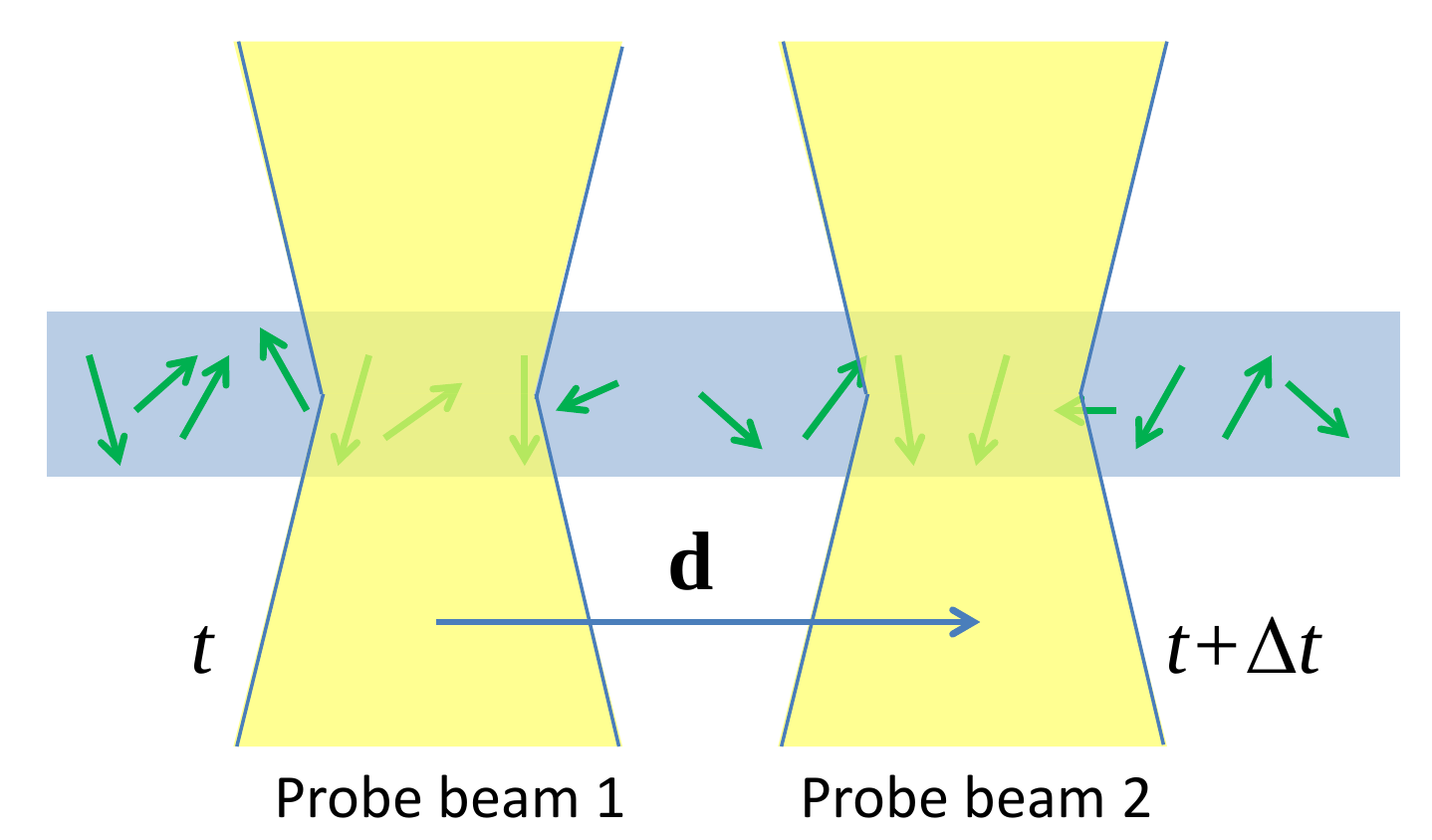}
\centering \caption{Two-beam spin noise spectroscopy.
Correlations between time-shifted intrinsic spin fluctuations at different space locations provide
information about spin transport at equilibrium. Here, $\mathbf{d}$ is the displacement of Beam 2 with respect to Beam 1.}
\label{fig1}
\end{figure}

We suggest a {\it two-beam spin noise spectroscopy} measuring both time and {\it space} correlations of spin fluctuations. Fig. \ref{fig1} presents a possible experimental setup involving two laser beams (that may or may not overlap) separated by a distance $d$. This setup captures the propagation of spin fluctuations between different space locations providing a unique information regarding spin transport at equilibrium. To the best of our knowledge, there are no other available experimental approaches to obtain such information.

In our theory, we assume that at the initial moment of time $t=0$ (that can be arbitrarily selected) the vector of spin polarization density is given by a vector of continuous random variables $\mathbf{S}(\mathbf{r},0)=\boldsymbol{\xi}(\mathbf{r})$ such that $\langle  \xi_i(\mathbf{r}) \rangle =0$ and $\langle \xi_i(\mathbf{r}) \xi_j(\mathbf{r'}) \rangle=\lambda \delta(\mathbf{r}-\mathbf{r'})\delta_{ij}$, where $\langle .. \rangle$ denotes averaging over different realizations, $i,j=x,y,z$, and $\lambda$ is a parameter describing the strength of spin fluctuations. Using statistical considerations~\cite{Reif65a}, one can find that $\lambda=n/4$, where $n$ is 2D electron density. For a given realization of initial spin polarization density, the spin polarization at $t>0$ can be written using a Green's function $G_{ij}$ of spin diffusion equation
\begin{equation}
S_i(\mathbf{r},t)=\int\limits_A G_{ij}(\mathbf{r},t;\mathbf{r'},0)S_j(\mathbf{r'},0)d\mathbf{r'}+\chi_i(\mathbf{r},t), \label{eq0}
\end{equation}
where $A$ is the sample area, $\chi_i(\mathbf{r},t)$ is a stochastic function describing the spin fluctuations created in the system at $t>0$ so that $\chi_i(\mathbf{r},0)=0$, and  $d\mathbf{r'}=dx'dy'$ (in 2D case). Next, we note that each beam in the optical setup shown in Fig. \ref{fig1} averages the space distribution of the Faraday rotation angle
$\theta (\mathbf{r},t)=\kappa S_z(\mathbf{r},t)$ according to
\begin{equation}
\bar{\theta}_m(t) =\frac{1}{P_0} \int\limits_A I_m(\mathbf{r})\theta (\mathbf{r},t)\textnormal{d}\mathbf{r}= \frac{\kappa}{P_0}\int\limits_A I_m(\mathbf{r})S_z (\mathbf{r},t)\textnormal{d}\mathbf{r}, \label{av_theta}
\end{equation}
$\kappa$ is a constant that couples $z$-component of spin polarization density with a local value of Faraday rotation angle, $P_0$ is the integrated laser beam intensity (power), and $I_m(\mathbf{r})$ is the space distribution of $m$-th beam intensity, where $m=1,2$.

In the case of two-beam spin noise spectroscopy (we assume that two beams with the same incident polarization and intensity are shifted by $\mathbf{d}$ as shown in Fig. \ref{fig1}), the experimentally determined correlation function is
\begin{equation}
R(t)=\langle \bar{\theta}_1(0) \bar{\theta}_2(t)\rangle. \label{R}
\end{equation}
Using Eqs. (\ref{eq0})-(\ref{R}), independence of $\boldsymbol{\chi}(\mathbf{r},t)$ from the initial spin density distribution $\boldsymbol{\xi}(\mathbf{r})$, and the expression $\langle \xi_i(\mathbf{r}) \xi_j(\mathbf{r'}) \rangle=\lambda \delta(\mathbf{r}-\mathbf{r'})\delta_{ij}$ (introduced above Eq. (\ref{eq0})), we find the general equation determining the second order spin noise correlation function:
\begin{equation}
R( t)=\frac{\lambda\kappa^2}{P_0^2}\int\limits_A \int\limits_A I(\mathbf{r}) I(\mathbf{r'}-\mathbf{d}) G_{zz}(\mathbf{r},t;\mathbf{r'}, 0) \textnormal{d}\mathbf{r}\textnormal{d}\mathbf{r'}. \label{R_final}
\end{equation}
The Fourier transform of $R(t)$ with respect to $t$ is the noise power spectrum
\begin{equation}
S(\omega)=\int\limits_0^\infty R(t)\cos(\omega t)\textnormal{d}t. \label{S(f)}
\end{equation}
We emphasize that although Eq. (\ref{R_final}) contains only $zz$ component of the Green function, the latter (as a solution of a system of spin diffusion equations) incorporates both transverse and longitudinal dynamics of spin polarization.

Eqs. (\ref{R_final}) and (\ref{S(f)})  are quite general and can be applied to many different experimental situations. Here, we consider a specific experimental system -- a semiconductor nanowire (made of, e.g., GaAs) with Bychkov-Rashba~\cite{Bychkov84a} spin-orbit interaction (SOI) described by $H_R=-\alpha\sigma_y\hat p$, where $\alpha$ is the spin-orbit coupling constant, $\sigma_y$ is the Pauli-matrix, and $\hat p$ is the $x$ component of the electron momentum operator. It is assumed that the wire diameter is much smaller than the spin diffusion length but much larger than the electron De Broglie wavelength. The strength of spin-orbit coupling is assumed to be smaller compared to the Fermi energy. The temperature is assumed to be in the range of 4-30K to allow for sufficiently long spin relaxation and suppression of purely quantum localization effects, so that we can describe the effective one-dimensional electron transport using coarse-grained Drude-model parameters, such as electron mobility, charge and spin diffusion length, etc., which is a standard approach to describe transport characteristics in this regime. Overall, our model is very close to that considered in Ref. \onlinecite{Glazov2011a}, however, our setup includes two probes instead of one~\cite{Glazov2011a}.

Moreover, we assume that the wire is placed in an in-plane magnetic field perpendicular to the wire axis (to shift a peak in the noise power spectrum from zero frequency) and, in addition to the D'yakonov-Perel' (DP) spin relaxation channel, there are additional spin relaxation mechanisms such as the Elliot-Yaffet mechanism~\cite{Yafet53a,Elliott54a} and/or electron spin relaxation on nuclear spins~\cite{Pershin2003a}.
Assuming a Gaussian distribution of the incident laser beam intensities along the $x$-direction, namely, $I(x)\propto \exp(-x^2/(2R_0^2))$, where $R_0$ is the beam radius, and using the Green's function of one-dimensional spin diffusion equation,
\begin{equation}
G_{zz}(x,t;x',0)=\frac{1}{\sqrt{4\pi D t}}e^{-\frac{(x-x')^2}{4Dt}-\gamma t}\cos\left(\eta (x-x')-\omega_Lt \right), \label{green_func}
\end{equation}
where $D$ is the diffusion coefficient, $\eta$ is the spin precession angle per unit length~\cite{pershin10a} (due to the Bychkov-Rashba SOI), $\gamma$ is the spin relaxation rate due to additional to the DP spin relaxation channels, and $\omega_L$ is the Larmor frequency, we find (with a help of Eq. (\ref{R_final})) the noise correlation function
\begin{equation}
R(t) \propto \frac{\cos \left(\omega_L t+\frac{d\eta D t}{Dt+R_0^2} \right)}{\sqrt{Dt+R_0^2}} e^{-\frac{R_0^2\eta^2Dt+d^2/4}{Dt+R_0^2}-\gamma t}. \label{res1}
\end{equation}

\begin{figure}[tb]
\centering\includegraphics[width=0.9\linewidth]{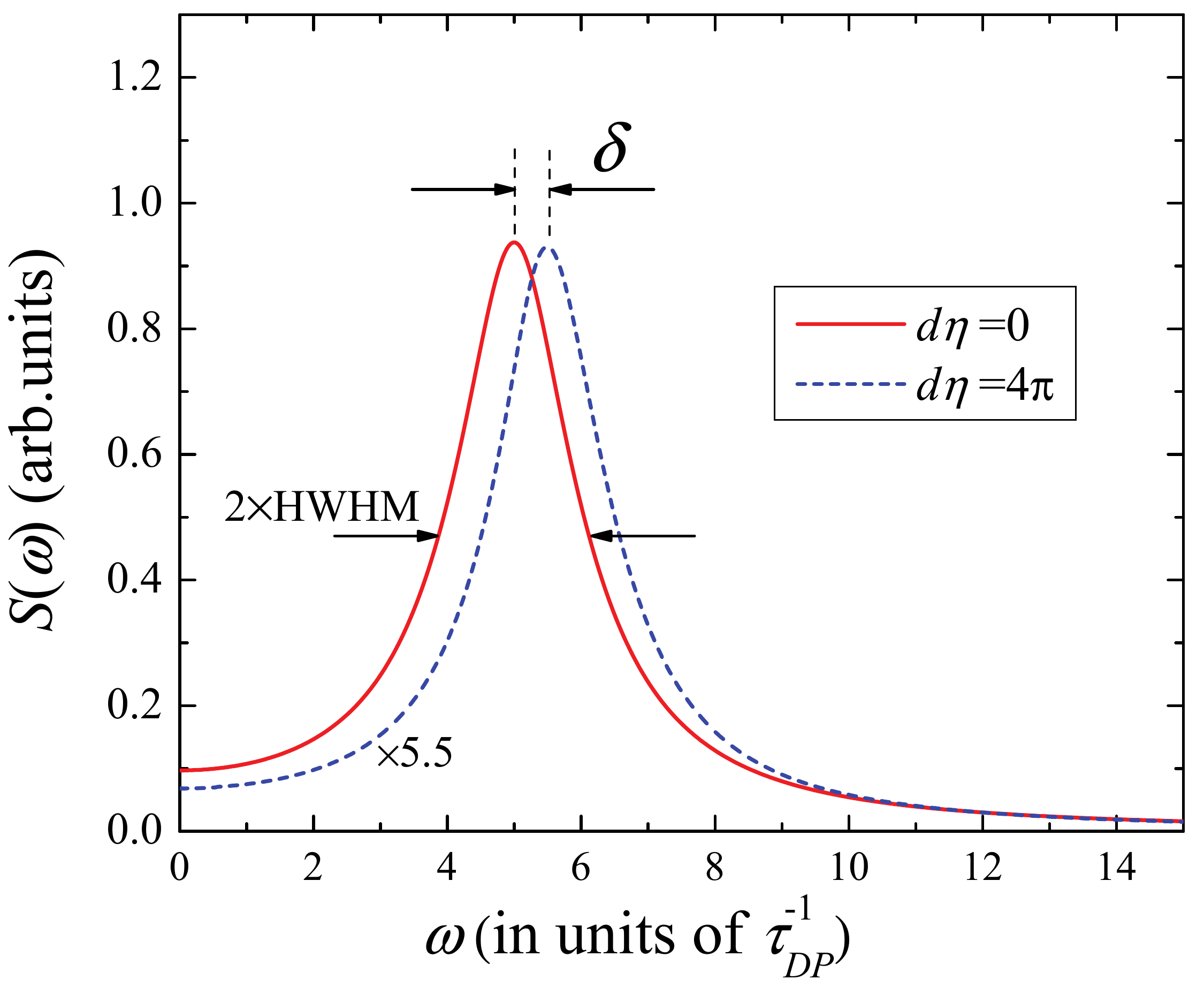}
\centering \caption{\label{fig2} Spin noise power spectrum in one-dimensional wires for different values of the beams' separation $d$. The direction of the peak's displacement is determined by the direction of the displacement of the second beam with respect to the first one along the wire. These curves were obtained using the parameter values $R_0\eta =1.5\pi$, $\gamma=0.2/ \tau_{DP}$, and $\omega_L=5/\tau_{DP}$.}
\end{figure}

Unfortunately, the Fourier transform of Eq. (\ref{res1}) can not be explicitly obtained. It is possible, however, to derive a closed analytical expression in a limiting case when the short times provide the main contribution to the Fourier transform of Eq. (\ref{res1}). Strictly speaking, this limiting case is realized if $R_0 | \eta | \gg 1$ and $d/(R_0^2 | \eta |) \ll 1$.
Neglecting $Dt$ compared to $R_0^2$ in Eq. (\ref{res1}), we rewrite this equation as
\begin{equation}
R(t) \propto \cos \left[ \left(\omega_L +\frac{d\eta D }{R_0^2}\right) t \right] e^{-(\gamma+\eta^2D)t}e^{-\frac{d^2}{4R_0^2}}. \label{res2}
\end{equation}
The Fourier transform of Eq. (\ref{res2}) gives the spin noise power spectrum
\begin{eqnarray}
 S(\omega )\propto
 \frac{\tau_s}{1+\tau_s^2(\omega_L+\delta-\omega )^2}e^{-\frac{d^2}{4R_0^2}}, \label{lorentz}
 \end{eqnarray}
where  $\tau_s^{-1}=\gamma+\tau_{DP}^{-1}$ is the total spin relaxation rate, $\tau_{DP}=(\eta^2D)^{-1}$ is the D'yakonov-Perel' spin relaxation time~\cite{Dyakonov72a,Dyakonov86a}, and $\delta=d\eta D/(R_0^2)$ is the peak shift from its position at $d=0$. According to Eq. (\ref{lorentz}), the spin noise spectrum shows a Lorentzian peak centered at $\omega'=\omega_L +\delta$; the resonance width is determined by the combined (due to all possible spin relaxation processes) spin relaxation time $\tau_s$. Importantly, the resonance is shifted from $\omega_L$ by $\delta=d\eta D /R_0^2$ depending on the strength of spin-orbit interaction (through the parameter $\eta$) and the beams' separation $d$. We emphasize that this shift of the resonance is an entirely new result that so far was not anticipated in the literature. We understand the resonance shift as follows. Depending on the direction of electron diffusion (in the positive or negative $x$ direction), the electron spins observed by the first beam acquire a spin rotation in the clockwise or counterclockwise direction (due to the spin-orbit interaction) by the moment of time when some of them are tested by the second beam in its own region of space. This spin rotation angle is added or subtracted from the Larmor precession angle causing a positive or negative peak shift.

\begin{figure}[tb]
\centering\includegraphics[width=0.9\linewidth]{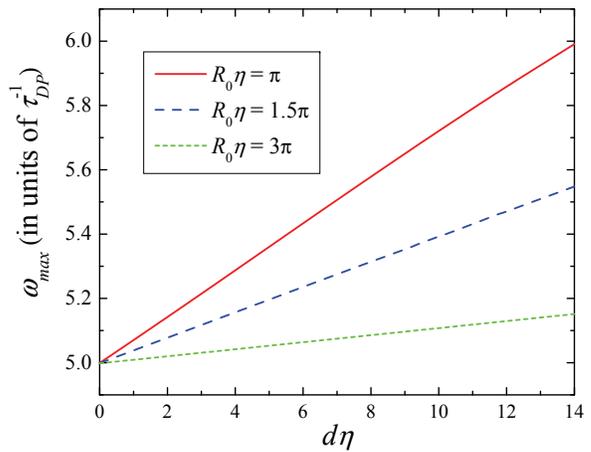}
\centering \caption{\label{fig3} Position of the peak in the spin noise power spectrum (Fig. \ref{fig2}) as a function of beams' separation $d$. These curves were obtained using the parameter values  $\gamma=0.2/ \tau_{DP}$ and $\omega_L=5/\tau_{DP}$.}
\end{figure}

\begin{figure}[tb]
\centering\includegraphics[width=0.9\linewidth]{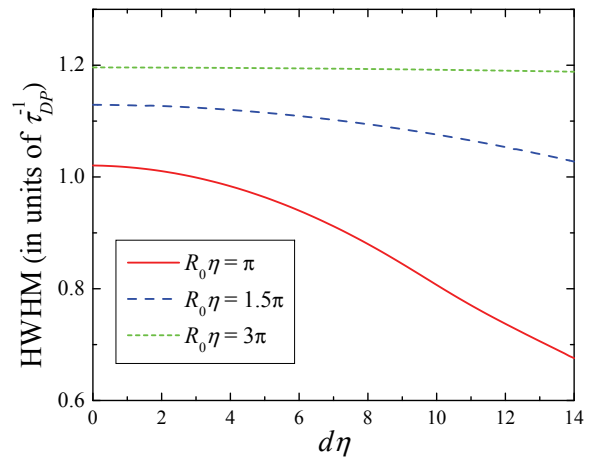}
\centering \caption{\label{fig4} Half width at half-maximum (HWHM) for the peak in the spin noise power spectrum (Fig. \ref{fig2}) as a function of beams' separation $d$. These curves were obtained using the parameter values  $\gamma=0.2/ \tau_{DP}$ and $\omega_L=5/\tau_{DP}$.}
\end{figure}

Fig. \ref{fig2} represents the spin noise power spectrum obtained as the Fourier transform of Eq. (\ref{res1}) for two values of $d$. Clearly, the peaks are of Lorentzian shape and $d\eta =4\pi$ peak is shifted by $\delta$ with respect to $d\eta =0$ peak. While the half width at half-maximum (HWHM) gives information about the possible spin relaxation channels, the value of the peak shift $\delta$ allows finding $\eta$, so that the strength of the DP spin relaxation and its role in overall spin relaxation can be determined.

Interestingly, additional effects can be observed if we consider a non-equilibrium situation e.g. by applying a weak external electric field $E$ along the wire. The
diffusion part of the propagator in (\ref{green_func}) is then modified to $\frac{1}{\sqrt{4\pi D t}} {\rm exp} \left( -(x-x'-\mu Et)^2/4Dt \right)$,
where $\mu$ is the electron mobility.
By neglecting $\sim E^2$ terms in corresponding
$R(t)$ function, the integration over the beam intensities gives us
\begin{equation}
S(\omega;E)\propto  \frac{\tau^{-1}_s+\frac{d\mu E}{2R_0^2}}{(\tau^{-1}_s+\frac{d\mu E}{2R_0^2})^2+(\omega_L+\delta-\eta \mu E-\omega)^2}e^{-\frac{d^2}{4R_0^2}},
\label{se}
\end{equation}
According to this result, the effect of the electric field is twofold. First, it renormalizes the effective correlation time  $\tau_s$ because spins moving along the electric field have a better chance to travel the distance $d$ between the two beams before they relax, than spins moving against the electric field. This effect should be observable as the change of the peak amplitude as a function of the electric field. Perhaps more importantly, the electric field renormalizes the position of the peak maximum: $\delta \rightarrow \delta - \eta \mu E$. This shift  can be used as an independent probe of the spin orbit coupling $\eta$. It emerges because the nonzero average drift velocity $ \mu E$, via  the spin orbit coupling, creates on average an effective transverse Zeeman-like field acting on electron spins with the corresponding Larmor frequency $\approx\eta \mu E$. This effect is not specific for quasi-1D geometry. It should emerge in any material with a linear in momentum spin-orbit coupling, such as the Rashba and the Dresslhaus couplings in 2D electron systems and strained 3D semiconductors. For example, taking parameters from Rashba electron system \cite{rashba-prl}: $\mu=\sigma_{xx}/(en)$, where $\sigma_{xx}= 1.8 \cdot 10^3\Omega^{-1}$m$^{-1}$ is the conductivity, and $\alpha h=1.5\cdot 10^{-13}$eV$\cdot$m, we find that an electric field  $E\sim400$V/cm  shifts the spin noise power peak by $\sim 100$MHz, which should be clearly observable  in GaAs.

Finally, we would like to consider $\delta$ and HWHM beyond the limiting case discussed above. For this purpose, we perform exact calculations of noise power spectrum (based on Eq. (\ref{res1})) and extract $\delta$ and HWHM from these calculations. Figs. \ref{fig3}, \ref{fig4} present some of our results. We note that while the peak shift scales almost linearly in the range of selected parameter values, the peak width demonstrates a significant variation at the smallest values $R_0 \eta$. Consequently, in the latter case the spin relaxation time should be extracted using the Fourier transform of the exact spin noise correlation function.

In conclusion, as the peak shift and width can be easily determined from experimental measurements, we anticipate that the suggested approach of {\it two-beam spin noise spectroscopy} will be extremely useful in the area of semiconductors and emergent materials. Possible applications of the two-beam spin noise spectroscopy include the possibility to separate contributions of conduction and localized electrons to the spin noise, to identify the role of different spin relaxation mechanisms contributing to the process of electron spin relaxation in a particular material, and to measure spin orbit coupling anisotropy. Additionally, the direction of in-plane magnetic field can be employed as an additional control parameter. We expect that it could be used to separate the contributions from the Rashba and Dresselhaus SOIs. Indeed, in the nanowire geometry, the Rashba and Dresselhaus\cite{Schliemann03a} SOIs cause the electron spin precessions in orthogonal directions that could be independently accessed selecting the direction of the magnetic field. However, this case is out of scope of this paper and is the subject for the future work.

{\it Acknowledgments.} This work has been partially supported by the University of South Carolina ASPIRE grant 13070-12-29502. The work at
LANL was carried out under the auspices of the National Nuclear
Security Administration of the U.S. Department of Energy at Los
Alamos National Laboratory under Contract No. DE-AC52-06NA25396.

\bibliography{spin}

\begin{thebibliography}{23}%
\makeatletter
\providecommand \@ifxundefined [1]{%
 \@ifx{#1\undefined}
}%
\providecommand \@ifnum [1]{%
 \ifnum #1\expandafter \@firstoftwo
 \else \expandafter \@secondoftwo
 \fi
}%
\providecommand \@ifx [1]{%
 \ifx #1\expandafter \@firstoftwo
 \else \expandafter \@secondoftwo
 \fi
}%
\providecommand \natexlab [1]{#1}%
\providecommand \enquote  [1]{``#1''}%
\providecommand \bibnamefont  [1]{#1}%
\providecommand \bibfnamefont [1]{#1}%
\providecommand \citenamefont [1]{#1}%
\providecommand \href@noop [0]{\@secondoftwo}%
\providecommand \href [0]{\begingroup \@sanitize@url \@href}%
\providecommand \@href[1]{\@@startlink{#1}\@@href}%
\providecommand \@@href[1]{\endgroup#1\@@endlink}%
\providecommand \@sanitize@url [0]{\catcode `\\12\catcode `\$12\catcode
  `\&12\catcode `\#12\catcode `\^12\catcode `\_12\catcode `\%12\relax}%
\providecommand \@@startlink[1]{}%
\providecommand \@@endlink[0]{}%
\providecommand \url  [0]{\begingroup\@sanitize@url \@url }%
\providecommand \@url [1]{\endgroup\@href {#1}{\urlprefix }}%
\providecommand \urlprefix  [0]{URL }%
\providecommand \Eprint [0]{\href }%
\providecommand \doibase [0]{http://dx.doi.org/}%
\providecommand \selectlanguage [0]{\@gobble}%
\providecommand \bibinfo  [0]{\@secondoftwo}%
\providecommand \bibfield  [0]{\@secondoftwo}%
\providecommand \translation [1]{[#1]}%
\providecommand \BibitemOpen [0]{}%
\providecommand \bibitemStop [0]{}%
\providecommand \bibitemNoStop [0]{.\EOS\space}%
\providecommand \EOS [0]{\spacefactor3000\relax}%
\providecommand \BibitemShut  [1]{\csname bibitem#1\endcsname}%
\let\auto@bib@innerbib\@empty
\bibitem [{\citenamefont {Zutic}, \citenamefont {Fabian},\ and\ \citenamefont
  {Das~Sarma}(2004)}]{Zutic04a}%
  \BibitemOpen
  \bibfield  {author} {\bibinfo {author} {\bibfnamefont {I.}~\bibnamefont
  {Zutic}}, \bibinfo {author} {\bibfnamefont {J.}~\bibnamefont {Fabian}}, \
  and\ \bibinfo {author} {\bibfnamefont {S.}~\bibnamefont {Das~Sarma}},\
  }\href@noop {} {\bibfield  {journal} {\bibinfo  {journal} {Rev. Mod. Phys.}\
  }\textbf {\bibinfo {volume} {76}},\ \bibinfo {pages} {323} (\bibinfo {year}
  {2004})}\BibitemShut {NoStop}%
\bibitem [{\citenamefont {Bandyopadhyay}\ and\ \citenamefont
  {Cahay}(2008)}]{Bandyopadhyay08a}%
  \BibitemOpen
  \bibfield  {author} {\bibinfo {author} {\bibfnamefont {S.}~\bibnamefont
  {Bandyopadhyay}}\ and\ \bibinfo {author} {\bibfnamefont {M.}~\bibnamefont
  {Cahay}},\ }\href@noop {} {\emph {\bibinfo {title} {Introduction to
  Spintronics}}}\ (\bibinfo  {publisher} {CRC Press},\ \bibinfo {year}
  {2008})\BibitemShut {NoStop}%
\bibitem [{\citenamefont {Young}\ \emph {et~al.}(2002)\citenamefont {Young},
  \citenamefont {Gupta}, \citenamefont {Johnston-Halperin}, \citenamefont
  {Epstein}, \citenamefont {Kato},\ and\ \citenamefont
  {Awschalom}}]{Young2002a}%
  \BibitemOpen
  \bibfield  {author} {\bibinfo {author} {\bibfnamefont {D.~K.}\ \bibnamefont
  {Young}}, \bibinfo {author} {\bibfnamefont {J.~A.}\ \bibnamefont {Gupta}},
  \bibinfo {author} {\bibfnamefont {E.}~\bibnamefont {Johnston-Halperin}},
  \bibinfo {author} {\bibfnamefont {R.}~\bibnamefont {Epstein}}, \bibinfo
  {author} {\bibfnamefont {Y.}~\bibnamefont {Kato}}, \ and\ \bibinfo {author}
  {\bibfnamefont {D.~D.}\ \bibnamefont {Awschalom}},\ }\href@noop {} {\bibfield
   {journal} {\bibinfo  {journal} {Semiconductor Science and Technology}\
  }\textbf {\bibinfo {volume} {17}},\ \bibinfo {pages} {275} (\bibinfo {year}
  {2002})}\BibitemShut {NoStop}%
\bibitem [{\citenamefont {M\"uller}\ \emph {et~al.}(2008)\citenamefont
  {M\"uller}, \citenamefont {R\"omer}, \citenamefont {Schuh}, \citenamefont
  {Wegscheider}, \citenamefont {H\"ubner},\ and\ \citenamefont
  {Oestreich}}]{Muller08a}%
  \BibitemOpen
  \bibfield  {author} {\bibinfo {author} {\bibfnamefont {G.~M.}\ \bibnamefont
  {M\"uller}}, \bibinfo {author} {\bibfnamefont {M.}~\bibnamefont {R\"omer}},
  \bibinfo {author} {\bibfnamefont {D.}~\bibnamefont {Schuh}}, \bibinfo
  {author} {\bibfnamefont {W.}~\bibnamefont {Wegscheider}}, \bibinfo {author}
  {\bibfnamefont {J.}~\bibnamefont {H\"ubner}}, \ and\ \bibinfo {author}
  {\bibfnamefont {M.}~\bibnamefont {Oestreich}},\ }\href@noop {} {\bibfield
  {journal} {\bibinfo  {journal} {Phys. Rev. Lett.}\ }\textbf {\bibinfo
  {volume} {101}},\ \bibinfo {pages} {206601} (\bibinfo {year}
  {2008})}\BibitemShut {NoStop}%
\bibitem [{\citenamefont {M\"uller}\ \emph {et~al.}(2010)\citenamefont
  {M\"uller}, \citenamefont {Oestreich}, \citenamefont {R\"omer},\ and\
  \citenamefont {H\"ubner}}]{Muller10a}%
  \BibitemOpen
  \bibfield  {author} {\bibinfo {author} {\bibfnamefont {G.~M.}\ \bibnamefont
  {M\"uller}}, \bibinfo {author} {\bibfnamefont {M.}~\bibnamefont {Oestreich}},
  \bibinfo {author} {\bibfnamefont {M.}~\bibnamefont {R\"omer}}, \ and\
  \bibinfo {author} {\bibfnamefont {J.}~\bibnamefont {H\"ubner}},\ }\href@noop
  {} {\bibfield  {journal} {\bibinfo  {journal} {Physica E: Low-dimensional
  Systems and Nanostructures}\ }\textbf {\bibinfo {volume} {43}},\ \bibinfo
  {pages} {569 } (\bibinfo {year} {2010})}\BibitemShut {NoStop}%
\bibitem [{\citenamefont {Oestreich}\ \emph {et~al.}(2005)\citenamefont
  {Oestreich}, \citenamefont {R\"omer}, \citenamefont {Haug},\ and\
  \citenamefont {H\"agele}}]{Oestreich2005a}%
  \BibitemOpen
  \bibfield  {author} {\bibinfo {author} {\bibfnamefont {M.}~\bibnamefont
  {Oestreich}}, \bibinfo {author} {\bibfnamefont {M.}~\bibnamefont {R\"omer}},
  \bibinfo {author} {\bibfnamefont {R.~J.}\ \bibnamefont {Haug}}, \ and\
  \bibinfo {author} {\bibfnamefont {D.}~\bibnamefont {H\"agele}},\ }\href@noop
  {} {\bibfield  {journal} {\bibinfo  {journal} {Phys. Rev. Lett.}\ }\textbf
  {\bibinfo {volume} {95}},\ \bibinfo {pages} {216603} (\bibinfo {year}
  {2005})}\BibitemShut {NoStop}%
\bibitem [{\citenamefont {Crooker}, \citenamefont {Cheng},\ and\ \citenamefont
  {Smith}(2009)}]{Crooker09a}%
  \BibitemOpen
  \bibfield  {author} {\bibinfo {author} {\bibfnamefont {S.~A.}\ \bibnamefont
  {Crooker}}, \bibinfo {author} {\bibfnamefont {L.}~\bibnamefont {Cheng}}, \
  and\ \bibinfo {author} {\bibfnamefont {D.~L.}\ \bibnamefont {Smith}},\
  }\href@noop {} {\bibfield  {journal} {\bibinfo  {journal} {Phys. Rev. B}\
  }\textbf {\bibinfo {volume} {79}},\ \bibinfo {pages} {035208} (\bibinfo
  {year} {2009})}\BibitemShut {NoStop}%
\bibitem [{\citenamefont {Glazov}\ and\ \citenamefont
  {Sherman}(2011)}]{Glazov2011a}%
  \BibitemOpen
  \bibfield  {author} {\bibinfo {author} {\bibfnamefont {M.~M.}\ \bibnamefont
  {Glazov}}\ and\ \bibinfo {author} {\bibfnamefont {E.~Y.}\ \bibnamefont
  {Sherman}},\ }\href@noop {} {\bibfield  {journal} {\bibinfo  {journal} {Phys.
  Rev. Lett.}\ }\textbf {\bibinfo {volume} {107}},\ \bibinfo {pages} {156602}
  (\bibinfo {year} {2011})}\BibitemShut {NoStop}%
\bibitem [{\citenamefont {Furis}\ \emph {et~al.}(2007)\citenamefont {Furis},
  \citenamefont {Smith}, \citenamefont {Kos}, \citenamefont {Garlid},
  \citenamefont {Reddy}, \citenamefont {Palmstrom}, \citenamefont {Crowell},\
  and\ \citenamefont {Crooker}}]{Furis07a}%
  \BibitemOpen
  \bibfield  {author} {\bibinfo {author} {\bibfnamefont {M.}~\bibnamefont
  {Furis}}, \bibinfo {author} {\bibfnamefont {D.~L.}\ \bibnamefont {Smith}},
  \bibinfo {author} {\bibfnamefont {S.}~\bibnamefont {Kos}}, \bibinfo {author}
  {\bibfnamefont {E.~S.}\ \bibnamefont {Garlid}}, \bibinfo {author}
  {\bibfnamefont {K.~S.~M.}\ \bibnamefont {Reddy}}, \bibinfo {author}
  {\bibfnamefont {C.~J.}\ \bibnamefont {Palmstrom}}, \bibinfo {author}
  {\bibfnamefont {P.~A.}\ \bibnamefont {Crowell}}, \ and\ \bibinfo {author}
  {\bibfnamefont {S.~A.}\ \bibnamefont {Crooker}},\ }\href@noop {} {\bibfield
  {journal} {\bibinfo  {journal} {New J. Phys.}\ }\textbf {\bibinfo {volume}
  {9}},\ \bibinfo {pages} {347} (\bibinfo {year} {2007})}\BibitemShut {NoStop}%
\bibitem [{\citenamefont {Mishchenko}(2003)}]{Mishchenko03a}%
  \BibitemOpen
  \bibfield  {author} {\bibinfo {author} {\bibfnamefont {E.~G.}\ \bibnamefont
  {Mishchenko}},\ }\href@noop {} {\bibfield  {journal} {\bibinfo  {journal}
  {Phys. Rev. B}\ }\textbf {\bibinfo {volume} {68}},\ \bibinfo {pages} {100409}
  (\bibinfo {year} {2003})}\BibitemShut {NoStop}%
\bibitem [{\citenamefont {Lamacraft}(2004)}]{Lamacraft04a}%
  \BibitemOpen
  \bibfield  {author} {\bibinfo {author} {\bibfnamefont {A.}~\bibnamefont
  {Lamacraft}},\ }\href@noop {} {\bibfield  {journal} {\bibinfo  {journal}
  {Phys. Rev. B}\ }\textbf {\bibinfo {volume} {69}},\ \bibinfo {pages} {081301}
  (\bibinfo {year} {2004})}\BibitemShut {NoStop}%
\bibitem [{\citenamefont {Egues}, \citenamefont {Burkard},\ and\ \citenamefont
  {Loss}(2002)}]{Egues02a}%
  \BibitemOpen
  \bibfield  {author} {\bibinfo {author} {\bibfnamefont {J.~C.}\ \bibnamefont
  {Egues}}, \bibinfo {author} {\bibfnamefont {G.}~\bibnamefont {Burkard}}, \
  and\ \bibinfo {author} {\bibfnamefont {D.}~\bibnamefont {Loss}},\ }\href@noop
  {} {\bibfield  {journal} {\bibinfo  {journal} {Phys. Rev. Lett.}\ }\textbf
  {\bibinfo {volume} {89}},\ \bibinfo {pages} {176401} (\bibinfo {year}
  {2002})}\BibitemShut {NoStop}%
\bibitem [{\citenamefont {Egues}\ \emph {et~al.}(2005)\citenamefont {Egues},
  \citenamefont {Burkard}, \citenamefont {Saraga}, \citenamefont {Schliemann},\
  and\ \citenamefont {Loss}}]{Egues05a}%
  \BibitemOpen
  \bibfield  {author} {\bibinfo {author} {\bibfnamefont {J.~C.}\ \bibnamefont
  {Egues}}, \bibinfo {author} {\bibfnamefont {G.}~\bibnamefont {Burkard}},
  \bibinfo {author} {\bibfnamefont {D.~S.}\ \bibnamefont {Saraga}}, \bibinfo
  {author} {\bibfnamefont {J.}~\bibnamefont {Schliemann}}, \ and\ \bibinfo
  {author} {\bibfnamefont {D.}~\bibnamefont {Loss}},\ }\href@noop {} {\bibfield
   {journal} {\bibinfo  {journal} {Phys. Rev. B}\ }\textbf {\bibinfo {volume}
  {72}},\ \bibinfo {pages} {235326} (\bibinfo {year} {2005})}\BibitemShut
  {NoStop}%
\bibitem [{\citenamefont {Reif}(1965)}]{Reif65a}%
  \BibitemOpen
  \bibfield  {author} {\bibinfo {author} {\bibfnamefont {F.}~\bibnamefont
  {Reif}},\ }\href@noop {} {\emph {\bibinfo {title} {Fundamentals of
  Statistical and Thermal Physics}}}\ (\bibinfo  {publisher} {McGraw-Hill},\
  \bibinfo {year} {1965})\BibitemShut {NoStop}%
\bibitem [{\citenamefont {Bychkov}\ and\ \citenamefont
  {Rashba}(1984)}]{Bychkov84a}%
  \BibitemOpen
  \bibfield  {author} {\bibinfo {author} {\bibfnamefont {Y.}~\bibnamefont
  {Bychkov}}\ and\ \bibinfo {author} {\bibfnamefont {E.}~\bibnamefont
  {Rashba}},\ }\href@noop {} {\bibfield  {journal} {\bibinfo  {journal} {{JETP}
  Lett.}\ }\textbf {\bibinfo {volume} {39}},\ \bibinfo {pages} {78} (\bibinfo
  {year} {1984})}\BibitemShut {NoStop}%
\bibitem [{\citenamefont {Yafet}(1953)}]{Yafet53a}%
  \BibitemOpen
  \bibfield  {author} {\bibinfo {author} {\bibfnamefont {Y.}~\bibnamefont
  {Yafet}},\ }\href@noop {} {\bibfield  {journal} {\bibinfo  {journal} {Solid
  State Phys.}\ }\textbf {\bibinfo {volume} {14}},\ \bibinfo {pages} {1}
  (\bibinfo {year} {1953})}\BibitemShut {NoStop}%
\bibitem [{\citenamefont {Elliott}(1954)}]{Elliott54a}%
  \BibitemOpen
  \bibfield  {author} {\bibinfo {author} {\bibfnamefont {R.~J.}\ \bibnamefont
  {Elliott}},\ }\href@noop {} {\bibfield  {journal} {\bibinfo  {journal} {Phys.
  Rev.}\ }\textbf {\bibinfo {volume} {96}},\ \bibinfo {pages} {266} (\bibinfo
  {year} {1954})}\BibitemShut {NoStop}%
\bibitem [{\citenamefont {Pershin}\ and\ \citenamefont
  {Privman}(2003)}]{Pershin2003a}%
  \BibitemOpen
  \bibfield  {author} {\bibinfo {author} {\bibfnamefont {Y.~V.}\ \bibnamefont
  {Pershin}}\ and\ \bibinfo {author} {\bibfnamefont {V.}~\bibnamefont
  {Privman}},\ }\href@noop {} {\bibfield  {journal} {\bibinfo  {journal} {Nano
  Letters}\ }\textbf {\bibinfo {volume} {3}},\ \bibinfo {pages} {695} (\bibinfo
  {year} {2003})}\BibitemShut {NoStop}%
\bibitem [{\citenamefont {Pershin}\ and\ \citenamefont
  {Slipko}(2010)}]{pershin10a}%
  \BibitemOpen
  \bibfield  {author} {\bibinfo {author} {\bibfnamefont {Y.~V.}\ \bibnamefont
  {Pershin}}\ and\ \bibinfo {author} {\bibfnamefont {V.~A.}\ \bibnamefont
  {Slipko}},\ }\href@noop {} {\bibfield  {journal} {\bibinfo  {journal} {Phys.
  Rev. B}\ }\textbf {\bibinfo {volume} {82}},\ \bibinfo {pages} {125325}
  (\bibinfo {year} {2010})}\BibitemShut {NoStop}%
\bibitem [{\citenamefont {Dyakonov}\ and\ \citenamefont
  {{Perel'}}(1972)}]{Dyakonov72a}%
  \BibitemOpen
  \bibfield  {author} {\bibinfo {author} {\bibfnamefont {M.~I.}\ \bibnamefont
  {Dyakonov}}\ and\ \bibinfo {author} {\bibfnamefont {V.~I.}\ \bibnamefont
  {{Perel'}}},\ }\href@noop {} {\bibfield  {journal} {\bibinfo  {journal} {Sov.
  Phys. Solid State}\ }\textbf {\bibinfo {volume} {13}},\ \bibinfo {pages}
  {3023} (\bibinfo {year} {1972})}\BibitemShut {NoStop}%
\bibitem [{\citenamefont {Dyakonov}\ and\ \citenamefont
  {Kachorovskii}(1986)}]{Dyakonov86a}%
  \BibitemOpen
  \bibfield  {author} {\bibinfo {author} {\bibfnamefont {M.~I.}\ \bibnamefont
  {Dyakonov}}\ and\ \bibinfo {author} {\bibfnamefont {V.~Y.}\ \bibnamefont
  {Kachorovskii}},\ }\href@noop {} {\bibfield  {journal} {\bibinfo  {journal}
  {Sov. Phys. Semicond.}\ }\textbf {\bibinfo {volume} {20}},\ \bibinfo {pages}
  {110} (\bibinfo {year} {1986})}\BibitemShut {NoStop}%
\bibitem [{\citenamefont {Engel}, \citenamefont {Halperin},\ and\ \citenamefont
  {Rashba}(2005)}]{rashba-prl}%
  \BibitemOpen
  \bibfield  {author} {\bibinfo {author} {\bibfnamefont {H.-A.}\ \bibnamefont
  {Engel}}, \bibinfo {author} {\bibfnamefont {B.~I.}\ \bibnamefont {Halperin}},
  \ and\ \bibinfo {author} {\bibfnamefont {E.~I.}\ \bibnamefont {Rashba}},\
  }\href@noop {} {\bibfield  {journal} {\bibinfo  {journal} {Phys. Rev. Lett.}\
  }\textbf {\bibinfo {volume} {95}},\ \bibinfo {pages} {166605} (\bibinfo
  {year} {2005})}\BibitemShut {NoStop}%
\bibitem [{\citenamefont {Schliemann}, \citenamefont {Egues},\ and\
  \citenamefont {Loss}(2003)}]{Schliemann03a}%
  \BibitemOpen
  \bibfield  {author} {\bibinfo {author} {\bibfnamefont {J.}~\bibnamefont
  {Schliemann}}, \bibinfo {author} {\bibfnamefont {J.~C.}\ \bibnamefont
  {Egues}}, \ and\ \bibinfo {author} {\bibfnamefont {D.}~\bibnamefont {Loss}},\
  }\href@noop {} {\bibfield  {journal} {\bibinfo  {journal} {Phys. Rev. Lett.}\
  }\textbf {\bibinfo {volume} {90}},\ \bibinfo {pages} {146801} (\bibinfo
  {year} {2003})}\BibitemShut {NoStop}%
\end{thebibliography}%

\end{document}